# New Organic thermally stable materials for optoelectronics devices - a linear spectroscopy study

Otilia Sanda Prelipceanu[1,3], Marius Prelipceanu[1,3], Ovidiu-Gelu Tudose[1,3], Bernd Grimm[2,3], Sigurd Schrader[1,3]

[1] University of Potsdam, Institute of Physics, Condensed Matter Physics, Am Neuen Palais 10, D-14469, Germany
[2] IDM, Institute of Thin Film Technology and Micro Sensorics, Kantstr.55, D-14513 Teltow, Germany
[3] University of Applied Sciences Wildau, Department of Engineering Physics, D- 15745 Wildau, Germany

Contact address: otiliap@rz.uni-potsdam.de

## 1. Introduction

Thermally stable polymers have attracted a lot of interest due to their potential use as the active component in electronic, optical and optoelectronic applications, such as light-emitting diodes, light emitting electrochemical cells, photodiodes, photovoltaic cells, field effect transistors, optocouplers and optically pumped lasers in solution and solid state.

Polymer-based structures are the focus of intensive investigations as mechanically and physically flexible, processible materials for large-area photoemitting and photosensitive devices. Their wide practical application is inhibited by present-day limitations in control over luminescent spectra, sensitivity and efficiency. We report results of our investigations into the use of thermal treatment of poly(p-phenylene vinylene) (PPV) films grown on a variety of substrates (quartz and glass). The samples studied had a thickness in the range 50 - 200 nm. Film thickness, morphology and structural properties were investigated by a range of techniques in particular: atomic force microscope - AFM, DEKTAK method, Ellipsometry and UV-VIS spectroscopy.

## 2. Experiment Part

Thin polymeric films are often used in the microelectronic industry, the development of optoelectronic applications. Homogeneous films with thickness varying from 50 – 200 nm are commonly prepared by spin coating. I this technique, polymers solution is dropped on the substrate surface (our case glass and quartz), which rotates at a given angular velocity during a give period of time. The film thickness is controlled by the concentration of the polymer is solution – 5% PPV in our experiment -, polymer molecular weight, spinning velocity and solvent evaporation rate. The polymers films are annealed at higher temperatures in vacuum and normal atmosphere and after this are investigated and results are compared. This work is concerned with the morphology of the thin films obtained from spin coating when different annealing method. The interactions between substrate, polymer and solvent were qualitatively correlated with the resulting surface morphology of spin coated films and treatment applied. We choose quartz and glass as substrates because this is transparent and easy for the spectroscopic investigations in transmission mode, and P-PPV dissolved in common solvents like toluene and chloroform. Moreover, the determination of the optical absorption and transmission, morphology and stability of the films are important for the development of electronic applications and waveguides[1].

Analytical grade toluene and chloroform were used to prepare the solutions at the polymer concentration 5 mg mL$^{-1}$. The P-PPV was dissolved in solvents, where no phase separation takes place. The chemical structure for PPV is schematically represented in Figure 1.

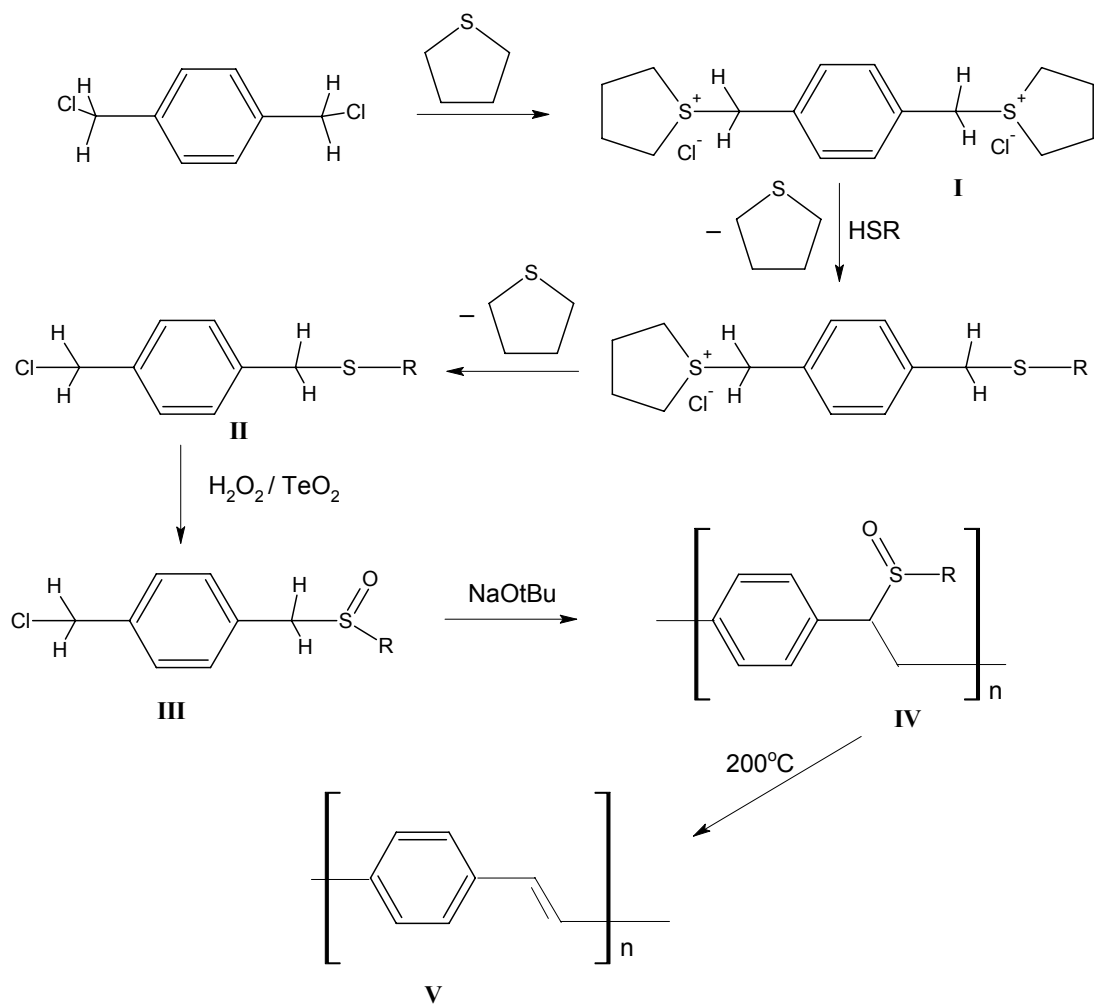

*Figure 1. Synthesis of PPV ( after C.J. Brabec, et al.)*

## 3. Methods and Results

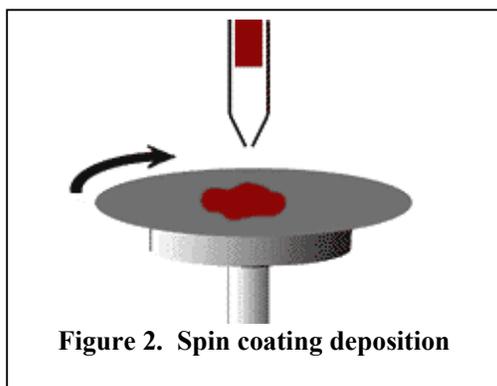

**Figure 2. Spin coating deposition**

*Spin coating* – The PPV films were prepared by spin coating on commercial quartz and glass substrates. The substrates dimensions of 1 cm x 2 cm were previously cleaned in standard manner and dried under a stream of $N_2$ [2]. All coatings were performed with the spinning velocity of 2000 rpm and the spinning time of 60 seconds.

*Ellipsometry* – The mean thickness and index of refraction (n) of the films were determined by means of ellipsometry in a Plasmos SD2000Automatic ellipsometer, Munich, Germany. The samples characteristics are shown in Table 1.

| Sample | Solvent | Thickness (nm) | Reflection index (n) |
|---|---|---|---|
| P-PPV on quartz (before annealing) | Toluene and Chloroform | 87 ± 5 | 1,3 ± 0.05 |
| PPV on Quartz (annealed in vacuum) | Toluene and Chloroform | 45 ± 5 | 2,590 ± 0.05 |
| PPV on quartz (annealed in normal atmosphere) | Toluene and Chloroform | 58 ± 5 | 2,567 ± 0.05 |
| PPV on quartz (after second annealing in vacuum) | Toluene and Chloroform | 44 ± 5 | 2,578 ± 0,05 |

**Table 1. Characteristic of PPV films obtained from spin coating. All measurements were performed at $24 \pm 2\ ^0 C$**

*Dektak measurements* – We measured and compared the morphology, thickness and aspect of films before and after treatment. For PPV films before annealing we obtained the thickness of 87 nm witch is shown in figure 3, and in figure 4 after annealing in vacuum we obtained 45 nm and figure 5 and 6 shown aspects layers before and after annealing. [3].

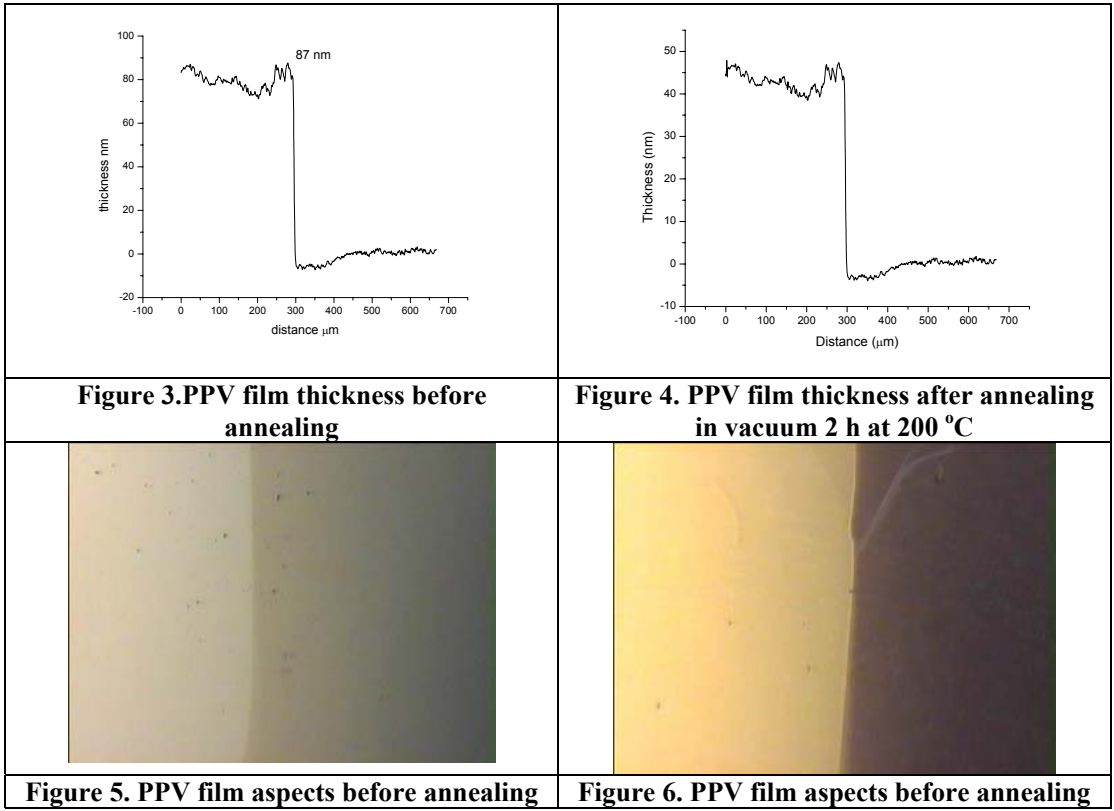

| Figure 3. PPV film thickness before annealing | Figure 4. PPV film thickness after annealing in vacuum 2 h at 200 °C |
| --- | --- |
| Figure 5. PPV film aspects before annealing | Figure 6. PPV film aspects before annealing |

*Atomic Force Microscopy* – Measurements were carried out with an instrument from Park Instrument Scientific (Sunnyvale, CA, USA) in non-contact mode in air at room temperature. All AFM images represent unfiltered original data and are displayed in color scale in figure 7, 8 and 9 [4].

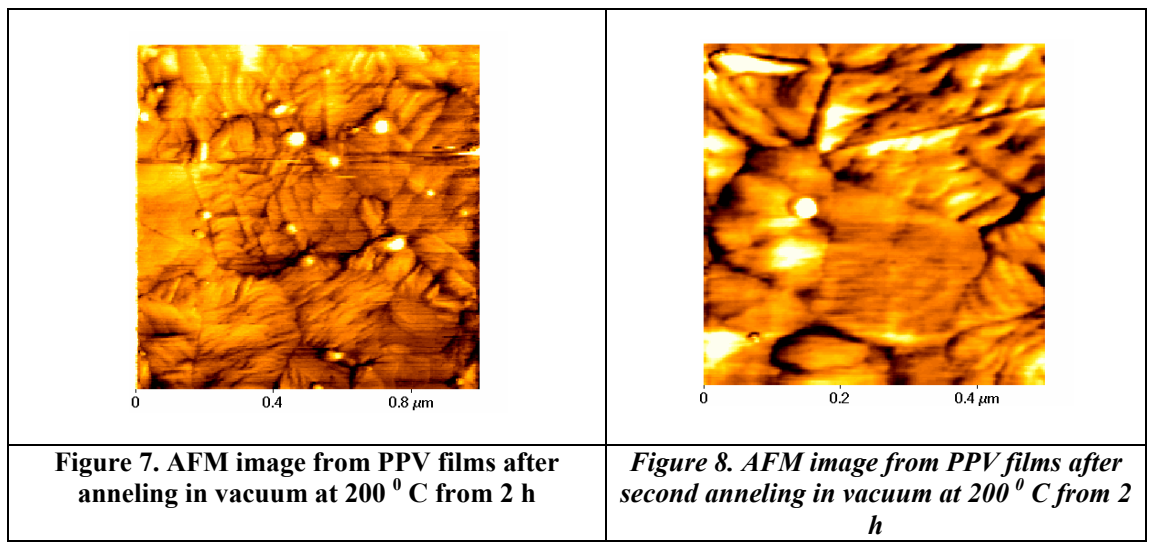

| Figure 7. AFM image from PPV films after anneling in vacuum at 200 [0] C from 2 h | *Figure 8. AFM image from PPV films after second anneling in vacuum at 200 [0] C from 2 h* |
| --- | --- |

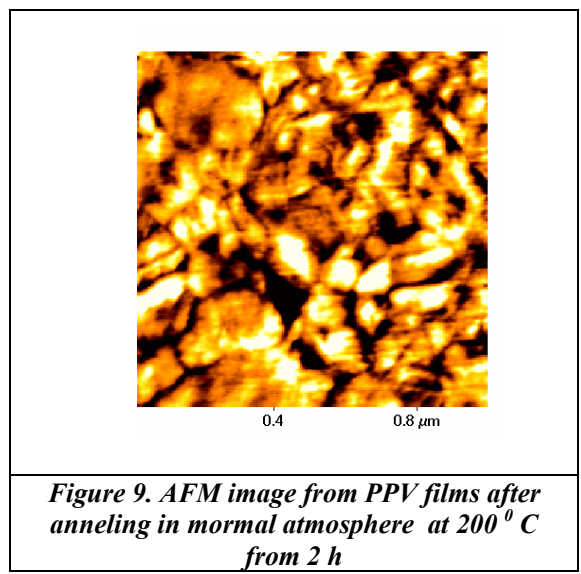
*Figure 9. AFM image from PPV films after anneling in mormal atmophere at 200 $^0$C from 2 h*

In figure 7 and 8 are shown the image of PPV films after first and second annealing in vacuum at 200 $^0$C for 2 hours. We can see not many changes between films, thichness were almost the same ( 45 nm respectively 44 nm) [5].

Figure 9 shows the surface structure of PPV films after annealing in normal atmophere at 200 $^0$C for 2 hours and the film structure are diferent, compare with films structure which were annealed in vacuum. For all images, films are continuous and smooth with a a root mean square (r.m.s) roughness of 2 – 3 nm, from the annealed in vacuum and 5-7 nm from the normal athmosphere annealed [6]. We get amorphous films in the both case. The main informations observed in Dektak measurements, AFM investigations and ellipsomentry are: the surface roughness of the films depend on the speed of heating, slow heat up raises the roughness, quick heat up leads to more smoth films. The same situation is meet in case of vacuum annealing and normal atmosphere annealing. More over, the thickness of the layers is reduced to about the haltf after annealing in the both case. The PPV layers are not orienteded in the both annealing method [6].

*UV/VIS measurements* - were made using the Perkin Elmer – UV/VIS Spectrometer Lambda 16. Spectra were acquired from 300 to 900 nm for optical excitation. Figure 9, 10 and 11 shows a set of absorption spectra of PPV films obtained from spin coating converted by heating under vacuum and normal atmosphere at 200$^o$C for 2 hours [7].

Spectra were normalized by dividing absorption spectrum of each individual sample by its absorption at the maximum. In this way relative changes within the spectrum and between the spectra are easily observed. One can notice differences in the position of the absorption maxima of PPV films prepared in different annealing method. The changes in the optical spectra of PPV films obtained from the precursors prepared in vacuum conditions and normal atmosphere condition are consistent with earlier observations in figure 10, 11 and 12 [8].

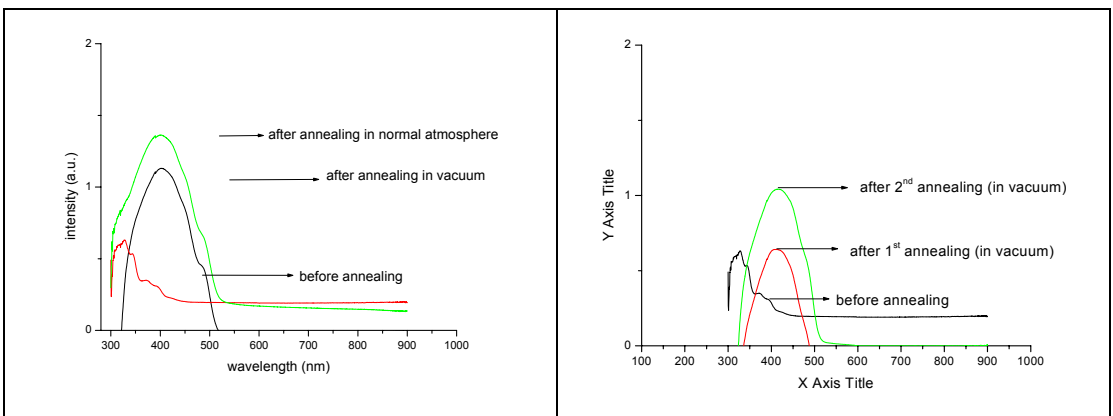

*Figure 10. UV-VIS spectra of PPV films obtained in vacuum conversion and normal atmosphere conversion*

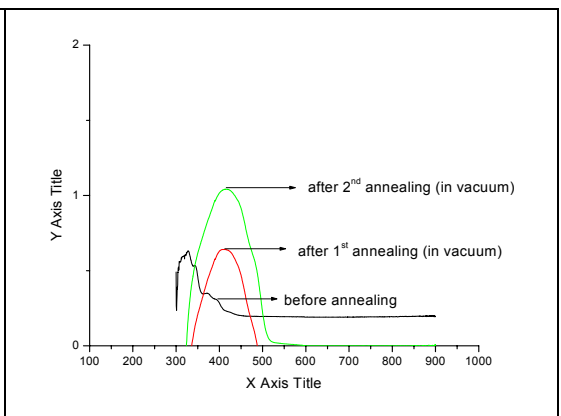

*Figure 11. UV-VIS spectra of PPV films obtained in vacuum conversion made several times*

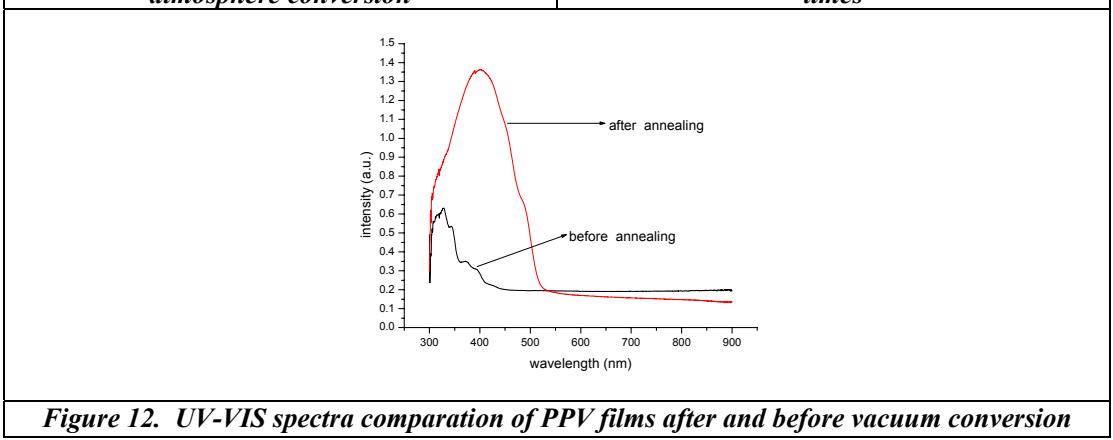

*Figure 12. UV-VIS spectra comparation of PPV films after and before vacuum conversion*

## 4. Conclusions

We summarize our findings as follows: (i) Annealing of PPV films causes ordering of polymer chains and, as a result, change in the luminescence intensity and spectra. (ii) spectral characteristics of the converted PPV-precursor strongly dependent on the preparation condition of the precursor (iii) the thickness of layers is reduced to about the half after annealing. (iv) The surface roughness of the films depends on the speed of heating: slow heat up raises the roughness; quick heat up leads to more smooth films. (v) PPV is thermally stable up to more than 500 $^0$C measured by TGA. (vi) We get amorphous films in spin coating deposition.

## 5. Acknowledgements

Financial support of the European Commission under contract number: **FP6 – 505478-1 ODEON - Project** and **RTN EUROFET – Project** is gratefully acknowledged.

## 6. References

[1] L. Bakueva, E.H. Sargent, R. Resendes, A. Bartole, I. Manners, J. Mater. Sci.: Mater. Electron. 12 (2001) 21.

[2] M. Pope, C.E. Swenberg, Electronic Processes in Organic Crystals and Polymers, Oxford Science Publications, Oxford, 1999.

[3] L. Bakueva, S. Musikhin, E.H. Sargent, A. Shik, 2001. MRS Fall Meeting, Boston, November 26–30, 2001 Book of Abstracts.

[4] D. Moses, A. Dogariu, A.J. Heeger, Synth. Met. 116 (2001) 19.

[5] B. Hu, F.E. Karaz, Chem. Phys. 227 (1998) 263.

[6] X.-R. Zeng, T.-M. Ko, J. Polym. Sci. B 35 (1997) 1993.

[7] C.E. Lee, C.-H. Jin, Synthet. Met. 117 (2001) 27.

[8] D.F.S. Petri, J.Braz.Chem.Soc. vol. 13, no 5, 695-699, 2002.